\documentclass[12pt]{article}
\usepackage{lscape}
\usepackage[usenames,dvipsnames]{xcolor}
\usepackage{fancyhdr}
\usepackage{verbatim,color,amssymb,epsfig}
\usepackage[usenames,dvipsnames]{xcolor}
\usepackage{subfigure}
\usepackage{alg}
\usepackage{fancyhdr}
\usepackage{amsmath}
\usepackage{enumitem}
\usepackage{epstopdf}
\usepackage[myheadings]{fullpage}
\usepackage{tikz}
\usepackage{color}
\usepackage{bm}

\def\boxit#1{\vbox{\hrule\hbox{\vrule\kern6pt
			\vbox{\kern6pt#1\kern6pt}\kern6pt\vrule}\hrule}}

\def\log{\hbox{log}}

\def\bse{\begin{eqnarray*}}
	\def\ese{\end{eqnarray*}}
\def\be{\begin{eqnarray}}
\def\ee{\end{eqnarray}}
\def\bq{\begin{equation}}
\def\eq{\end{equation}}
\def\bse{\begin{eqnarray*}}
	\def\ese{\end{eqnarray*}}

\def\b1e{{\mathbf e}}

\begin{document}

	\thispagestyle{empty} \baselineskip=28pt
	
	\begin{center}
		
		{\LARGE \textbf{An EM Algorithm for Estimating an Oral Reading Speed and Accuracy Model}}
		
	\end{center}
	
	\baselineskip=10pt
	
	\vskip 2mm

	\begin{center}
		Cornelis Potgieter\\
		Department of Statistical Science, Southern Methodist University, Dallas TX 75275, cpotgieter@smu.edu\\
		\hskip 5mm \\
		Akihito Kamata\\
		Simmons School of Education, \& Department of Psychology, Southern Methodist University, Dallas, TX 75206 \\
		\hskip 5mm\\
		Yusuf Kara\\
		Department of Educational Measurement and Evaluation, Anadolu University, Eskisehir, Turkey \\

	\end{center}

	\vskip 5mm
	\begin{center}
		{\Large \textbf{Abstract}}
	\end{center}

	This study proposes a two-part model that includes components for reading accuracy and
	reading speed. The speed component is a log-normal factor model, for which speed data are
	measured by reading time for each sentence being assessed. The accuracy
	component is a binomial-count factor model, where the accuracy data are measured
	by the number of correctly read words in each sentence. Both underlying
	latent components are assumed to be Gaussian in nature. In this paper, the
	theoretical properties of the proposed model are developed and an Monte Carlo EM
	algorithm for model fitting is outlined. The predictive power of the model
	is illustrated in a real data application.
	\vskip 10mm
	
	\baselineskip=12pt
	
\pagenumbering{arabic} \newlength{\gnat}

\section{Introduction}

As part of screening assessments of reading ability in schools, oral reading
fluency (ORF) is frequently assessed to identify students at-risk for poor
learning outcomes in order to help guide and inform instructional
decision-making for such students (e.g., Fuchs, 2004). In traditional ORF
test administration, a student is given one minute to read as many words as
possible in a grade-level text of approximately 250 words. During the test
administration, a trained assessor follows along and indicates on a scoring
protocol each word the student reads incorrectly. After one minute, the
total number of words correctly read is obtained. If the student finishes
reading the entire passage within one minute, the time took to read the
passage is also obtained. Then, they are transformed into a reporting score,
namely, the number of words correctly read per minute (wcpm).

Despite prevalent use and practical application of ORF measures, the current
standard assessment of ORF has considerable psychometric limitations which potentially
make ORF measures less reliable and valid. This paper is part of an effort to develop and establish a new ORF assessment system. The new ORF assessment system incorporates
centralized scoring based on recorded reading both by human assessor and
speech recognition engine. As a consequence, the assessment system collects
accuracy and time data at the word, sentence and passage levels.
Availabilities of word and sentence level data enables one to estimate
reading speed beyond the traditional wcpm scores. Thus, this study proposes
and evaluates a psychometric model to estimate reading speed by a latent
variable model that incorporates speed and accuracy jointly.

The proposed model is a modification of a speed-accuracy model proposed by
van der Linden (2007). van der Linden's model is a two-part model for speed and accuracy when taking a test. The speed component is a
log-normal factor model, for which speed data are measured by time taken (such as
second) to respond to each item being assessed. The accuracy component
is a 3-parameter logistic (3-PL) item response theory model, for which
accuracy data are measured by correct-incorrect item responses. In the present
study, the speed part of the model also follows a log-normal factor model as formulated by van der Linden, and the speed data are measured by the time it took to read each sentence. On
the other hand, this study utilizes a a two-component normal ogive binomial-count factor model, where the
accuracy data are measured by the number of correctly read words in each
sentence.

The approach to model fitting here is explicitly frequentist in nature, while the
model proposed by van der Linden (2007)\ was fit in a hierarchical Bayes
framework. Hierarchical Bayes is a common approach with these type of latent variable
models - see for example Fox \textit{et al.} (2007), Entink \textit{et al.} (2009) and van der Linden \textit{et al.} (2010). In this paper, we propose an EM\ algorithm for fitting the model to
avoid the computational complexity associated with direct maximization of
the likelihood function. The EM\ algorithm was first proposed by Dempster, Laird \& Rubin (1977) for performing maximum likelihood estimation in the presence of missing data; see also the recent monograph by McLachlan \& Krishnan (2007) and the references therein. The modeling approach adopted in this paper makes use of a variation of the EM algorithm known as the Monte
Carlo EM\ (MCEM) algorithm, details of which can be found in Wei \& Tanner (1990).

The outline of the paper is as follows. In Section 2, the
proposed model is outlined and some of its theoretical properties are
discussed. Section 3 discussed two approaches
to parameter estimation, a method of moments approach and maximum likelihood
estimation via the MCEM\ algorithm. Section 4 presents
simulation results comparing the two proposed methods. Finally, an analysis
of a real dataset is presented in Section 5. This
analysis includes an investigation of the out-of-sample predictive ability
of the proposed model. After concluding remarks in Section 6, the derivation of some technical results and the sampling
algorithm used for implementation of the MCEM\ method are outlined
in an appendix.

\section{Oral Reading Fluency Model}

So far, we have described the context of the model assuming that the unit of
analysis was a sentence. However, the unit can be a word, a sentence or a paragraph. 
Therefore, the generic term \textit{item} will be used hereafter to refer
to the unit of analysis.
Let $\mathbf{N}=\left( N_{1},\ldots
,N_{I}\right) $ denote the vector of total words per item where items are indexed $i=1,\ldots ,I$. For the $j^{th}$
individual, $j=1,\ldots ,n$, let $\mathbf{Y}_{j}=\left( Y_{1j},\ldots
,Y_{Ij}\right) $ denote the response vector containing number of words read
correctly for each of the $I$ items, and let $\mathbf{T}_{j}=\left(
T_{1j},\ldots ,T_{Ij}\right) $ denote the response vector of reading times for each of the $I$ items. Furthermore, let $\boldsymbol{\xi }_{j}=\left( \theta _{j},\tau _{j}\right) $ denote the pair of latent
variables with $\theta _{j}$ the reading accuracy factor and $\tau _{j}
$ the reading speed factor. In the model, it is assumed that, conditional on ability, words correct per item follows a Binomial distribution,
\begin{equation}
Y_{ij}|\theta _{j}\sim f\left( y_{ij};N_{i},p_{i}\left( \theta _{j}\right)
\right)
\label{Y_given_theta}
\end{equation}%
where $f$ denotes the binomial mass function with item size $N_{i}$ and success probability $p_{i}\left( \theta _{j}\right) =\Phi \left( a_{i}\left( \theta_{j}-b_{i}\right) \right) $. Here $a_{i}\in \mathbb{R} ^{+}$ and $b_{i}\in \mathbb{R} $ are interpreted as the discrimination and
difficulty of the $i^{th}$ item's differentiation of reading accuracy, respectively, and $\Phi \left( \cdot
\right) $ denotes the standard normal cumulative distribution function. Conditional on latent speed, a log-normal model is specified for response time, so that
\begin{equation}
\hbox{log}T_{ij}|\tau _{j}\sim \alpha _{i}\phi \left( \alpha _{i}\left(
t_{ij}-\beta _{i}+\tau _{j}\right) \right) 
\label{T_given_tau}
\end{equation}%
where $t_{ij}$ denotes the natural logarithm of the time it takes individual $j$ to read item $i$. Here $\alpha _{i}\in \mathbb{R} ^{+}$ and $\beta_{i}\in \mathbb{R} $ are interpreted as the discrimination
and intensity parameters of the $i^{th}$ item's differentiation of reading speed. Here, $\phi \left( \cdot
\right) $ denotes the standard normal density function. Combining (\ref{Y_given_theta}) and (\ref{T_given_tau}), the joint
distribution of response vector $\left( \mathbf{Y}_{j},\log \mathbf{T}%
_{j}\right) $ conditional on $\boldsymbol{\xi }_{j}$ is given by%
\begin{equation*}
f\left( \mathbf{y}_{j},\mathbf{t}_{j}|\boldsymbol{\xi }_{j}\right)
=\prod_{i=1}^{n}f\left( y_{ij};n_{i},p_{i}\left( \theta _{j}\right) \right)
\alpha _{i}\phi \left( \alpha _{i}\left(t_{ij} - \beta _{i} + \tau _{j}\right) \right) 
\end{equation*}%
where $\mathbf{y}_{j}=(y_{1j},\ldots,y_{Ij})$ and $\mathbf{t}_{j}=(t_{1j},\ldots,t_{Ij})$ denotes the observed counts and logarithmic times for the $j^{th}$ individual. Note that for
a given item, the words correct count and reading time are conditionally independent
given latent vector $\boldsymbol{\xi }_{j}$. It is assumed that this latent
vector $\bm{\xi }_{j}$ follows a bivariate normal distribution with mean vector $\bm{\mu }_{\bm{\xi }}=\left( \mu _{\theta },\mu _{\tau }\right)$
and covariance matrix%
\begin{equation*}
\bm{\Sigma }_{\bm{\xi }}=\left( 
\begin{array}{cc}
\sigma _{\theta }^{2} & \sigma _{\theta \tau } \\ 
\sigma _{\theta \tau } & \sigma _{\tau }^{2}%
\end{array}%
\right) .
\end{equation*}%
For model identifiability, it is necessary to impose constraints on these parameters and therefore it is assumed that $\mu _{\theta
}=\mu _{\tau }=0$ and that $\sigma _{\theta }^{2}=1$, while $\sigma _{\theta \tau }$ and $\sigma _{\tau }^{2}$ are free parameters.  No
distributional assumptions are made for the item-specific parameters $\left\{ \left(
a_{i},b_{i},\alpha _{i},\beta _{i}\right) \right\} _{i=1,\ldots ,I}$. This
could certainly be incorporated into the model and would result in a nonlinear random effects model. However, the model developed here does not
assume items are randomly drawn from some population of items. Rather, the
model is developed conditional on the collection of $I$ items used in assessing reading accuracy and speed. 

The unconditional distribution of response vector $\left( \mathbf{Y}_{j},\log \mathbf{T}%
_{j}\right) $ is given by%
\begin{equation}
f\left( \mathbf{y}_{j},\mathbf{t}_{j}\right) =\int_{\mathbb{R} ^{2}}\left[
\prod_{i=1}^{I}f\left( y_{ij};N_{i},p_{i}\left( \theta \right) \right)
\alpha _{i}\phi \left( \alpha _{i}\left( t_{ij} - \beta _{i} + \tau \right) \right) %
\right] \phi _{2}\left( \boldsymbol{\xi };\bm{\mu }_{\bm{\xi }},\mathbf{%
	\Sigma }_{\bm{\xi }}\right) d\bm{\xi }  \label{Joint Dist YT}
\end{equation}%
where the integral is taken over the real plane $\boldsymbol{\xi } = (\theta,\tau)\in \mathbb{R} ^{2}$ and where $\phi
_{2}\left( \mathbf{\cdot };\bm{\mu },\bm{\Sigma }\right) $ denotes
the bivariate normal density with mean $\bm{\mu }$ and covariance $%
\bm{\Sigma }$. 

It is not uncommon for datasets to contain missing values, as an individual may not read all items in the alloted time. Recording errors can also result in missing values. It is assumed that, for the $j^{th}$ individual and $i^{th}$ item, either both or neither the count $y_{ij}$ and log-time $t_{ij}$ are observed. When missing, the pair of measurements is assumed to be missing completely at random. Let $%
\mathcal{S}_{j}\subseteq \left\{ 1,2,\ldots ,I\right\} $ denote the set of
items for which count and time variables were observed for the $j^{th}$
individual. The joint distribution which takes missing values into account
can be written as%
\begin{equation}
f\left( \mathbf{y}_{j},\mathbf{t}_{j}|\mathcal{S}_{j}\right) =\int_{\mathbb{R} ^{2}}%
\left[ \prod_{i\in \mathcal{S}_{j}}f\left( y_{ij};N_{i},p_{i}\left( \theta
\right) \right) \alpha _{i}\phi \left( \alpha _{i}\left( t_{ij}-\beta _{i}+\tau
\right) \right) \right] \phi _{2}\left( \bm{\xi };\bm{\mu }_{\bm{%
		\xi }},\bm{\Sigma }_{\bm{\xi }}\right) d\bm{\xi }.
\label{Joint Dist YT with missing}
\end{equation}%
The integral form of this joint distribution makes direct maximization of
the likelihood function an untenable proposition. This, in part, speaks to
the popularity of Bayesian methods in estimating complex latent factor models that
are nonlinear in nature. One of the contributions the present paper makes to the literature is the development of an EM algorithm for maximizing a likelihood function based on (\ref{Joint Dist YT}). The EM algorithm, as well as a method of moments approach to parameter estimation, is discussed in the next section.

\section{Parameter Estimation}

\subsection{Method of Moments}

In this subsection, a method of moments (MOM) method is proposed for estimating the parameter vectors $\left\{ \left( a_{i},b_{i},\alpha _{i},\beta _{i}\right) \right\}
_{i=1,\ldots ,I}$ as well as the parameters $\sigma _{\tau }^{2}$ and $\sigma _{\theta \tau
}$. The MOM estimator, while interesting in their own right, also provide useful starting values for the EM algorithm which is outlined
in the next subsection. The moments of $\log T_{ij}$ follow from properties of the normal distribution and derivation is omitted. The moments of the count variables $Y_{ij}$ and the covariance between $Y_{ij}$ and $\log T_{ij}$ are derived in Appendix 7.1.

For the $i^{th}$ item, the word count $Y_{ij}$ has mean%
\begin{equation}
\mathrm{E}\left( Y_{ij}\right) =N_{i}\Phi \left( -\frac{a_{i}b_{i}}{\sqrt{%
		1+a_{i}^{2}}}\right)   \label{E_Y}
\end{equation}%
and variance%
\begin{eqnarray}
\mathrm{Var}\left( Y_{ij}\right)  &=&N_{i}^{2}\left[ \Phi _{2}\left( -\frac{%
	a_{i}b_{i}}{\sqrt{1+a_{i}^{2}}},-\frac{a_{i}b_{i}}{\sqrt{1+a_{i}^{2}}}%
\left\vert \rho =\frac{a_{i}^{2}}{1+a_{i}^{2}}\right. \right) -\Phi
^{2}\left( -\frac{a_{i}b_{i}}{\sqrt{1+a_{i}^{2}}}\right) \right] 
\label{Var_Y} \\
&&+N_{i}\left[ \Phi \left( -\frac{a_{i}b_{i}}{\sqrt{1+a_{i}^{2}}}\right)
-\Phi _{2}\left( -\frac{a_{i}b_{i}}{\sqrt{1+a_{i}^{2}}},-\frac{a_{i}b_{i}}{%
	\sqrt{1+a_{i}^{2}}}\left\vert \rho =\frac{a_{i}^{2}}{1+a_{i}^{2}}\right.
\right) \right]   \notag
\end{eqnarray}%
where $\Phi _{2}(\cdot,\cdot|\rho)$ denotes the bivariate normal cumulative distribution function with zero means, unit variances and correlation coefficient $\rho $. The model can accommodate both over- and under-dispersed count data, as the unconditional variance of $Y_{ij}$ can be either larger or smaller than the mean for appropriate choices of parameter values $a_{i}$ and $b_{i}$.

Next, consider the logarithm of reading time per item, $\log T_{ij}$. Using standard properties of the normal distribution, it follows that%
\begin{equation}
\mathrm{E}\left( \log T_{ij}\right) =\beta _{i}  \label{E_logT},
\end{equation}%
\begin{equation}
\mathrm{Var}\left( \log T_{ij}\right) =\sigma _{\tau }^{2}+\frac{1}{\alpha
	_{i}^{2}}  \label{Var_logT}
\end{equation}%
and%
\begin{equation}
\mathrm{Cov}\left( \log T_{ij},\log T_{i^{\prime }j}\right) =\sigma _{\tau
}^{2}  \label{Cov_logT}
\end{equation}%
where $i\not=i^{\prime }$. Finally, the covariance between the word count and the logarithm of reading time is

\begin{equation}
\mathrm{Cov}\left( Y_{ij},\hbox{log}T_{ij}\right) =-\sigma _{\tau \theta }%
\frac{n_{i}}{\sqrt{2\pi }}\left( \frac{a_{i}^{2}}{a_{i}^{2}+1}\right)
^{1/2}\exp \left( -\frac{1}{2}\frac{a_{i}^{2}b_{i}^{2}}{a_{i}^{2}+1}\right) .
\label{Cov_Y_logT}
\end{equation}

Method of moments estimators are found by replacing the population
moments in equations (\ref{E_Y}), (\ref{Var_Y}), (\ref{E_logT}), (\ref%
{Var_logT}), (\ref{Cov_logT}) and (\ref{Cov_Y_logT}) by their sample equivalents and then solving for the unknown parameters. This can sometimes be done in multiple ways, specifically when there are more moment equations than unknown parameters. Therefore, the estimators presented below are only one possible way of finding MOM estimators.

For the $i^{th}$ item, let $\bar{y}_i$ and $s_{y_i}^2$ denote the sample mean and variance of observed counts $y_{ij}$, calculated over the set of individuals with non-missing responses for the $i^{th}$ item,  $\{j:i \in \mathcal{S}_{j}\}$. Similarly, let $\bar{t}_i$ and $s_{t_i}^2$ denote the sample mean and variance of the $i^{th}$ item's log-times $t_{ij}$. Now, let $\hat{\rho}_{i}$ be the
correlation coefficient that solves estimating equation%
\begin{equation}
\Phi _{2}\left( \left. \Phi ^{-1}\left( \frac{\bar{y}_{i}}{N_{i}}\right)
,\Phi ^{-1}\left( \frac{\bar{y}_{i}}{N_{i}}\right) \right\vert \rho
_{i}\right) =\frac{s_{y_{i}}^{2}+\bar{y}_{i}\left( \bar{y}_{i}-1\right) }{%
	N_{i}\left( N_{i}-1\right) }.  \label{EstEq_rho}
\end{equation}%
Subsequently, estimators of $a_i$ and $b_i$ are given by
\begin{equation}
\hat{a}_{i}=\left( \frac{\hat{\rho}_{i}}{1-\hat{\rho}_{i}}\right) ^{1/2}
\label{ab_EstEq1}
\end{equation}%
and%
\begin{equation}
\hat{b}_{i}=-\frac{\left( 1+\hat{a}_{i}^{2}\right) ^{1/2}}{\hat{a}_{i}}\Phi
^{-1}\left( \frac{\bar{y}_{i}}{N_{i}}\right)   \label{ab_EstEq2}
\end{equation}%
for $i=1,\ldots ,I$. Now, let $s_{t_{i},t_{i^{\prime }}}$ denote the sample
covariance between the log-times of items $i$ and $i^{\prime }$, covariance calculated over the set $\{j:i,i^{\prime} \in \mathcal{S}_{j}\}$, i.e., individuals for which both items $i$ and $i^{\prime}$ are observed. An estimator of $\sigma^2_{\tau}$ is given by%
\begin{equation}
\hat{\sigma}_{\tau }^{2}=\frac{2}{I\left( I-1\right) }\sum_{i=1}^{I}%
\sum_{i^{\prime }=i+1}^{I}s_{t_{i},t_{i^{\prime }}} . \label{sigma2_tau}
\end{equation}%
Subsequently, define%
\begin{equation}
\hat{\alpha}_{i}=\frac{1}{\max \left[ 0,\left( s_{t_{i}}^{2}-\hat{\sigma}%
	_{\tau }^{2}\right) ^{1/2}\right] }  \label{alpha_est}
\end{equation}%
and%
\begin{equation}
\hat{\beta}_{i}=\bar{t}_{i}  \label{beta_est}
\end{equation}%
for $i=1,\ldots ,I$. The estimator in (\ref{alpha_est}) contains a finite-sample correction to ensure that $\hat{\alpha}_i\geq0$ for all $i$. Note that this moment-estimator of $\alpha_i$ can be infinite, but this doesn't pose a problem as $\mathrm{Var}(Y_{ij})$ in (\ref{Var_logT}) depends on the inverse of $\alpha_i$. Finally, let $s_{y_{i},t_{i}}$ denote the covariance
between word count and the log-time for the $i^{th}$ item
and define%
\begin{equation}
\hat{\sigma}_{\theta \tau }=-\frac{\left( 2\pi \right) ^{1/2}}{I}%
\sum_{i=1}^{I}\frac{s_{y_{i},t_{i}}}{n_{i}}\left( \frac{\hat{a}_{i}^{2}}{%
	\hat{a}_{i}^{2}+1}\right) ^{-1/2}\exp \left( \frac{1}{2}\frac{\hat{a}_{i}^{2}%
	\hat{b}_{i}^{2}}{\hat{a}_{i}^{2}+1}\right) .  \label{sigma_theta_tau_est}
\end{equation}%
Equations (\ref{ab_EstEq1}) through (\ref{sigma_theta_tau_est}) are the proposed MOM estimators of the model parameters. These MOM estimators have an advantage over the maximum likelihood estimators in
that they are fast and easy to calculate. However, the MOM estimators are 
typically less efficient than the maximum likelihood estimators when compared using root mean square error (RMSE) as a criterion. This will be illustrated in the simulation study section of this paper.

\subsection{Monte Carlo EM\ Algorithm}

The EM\ algorithm, originally proposed by Dempster et al. (1977), is a method of performing maximum likelihood estimation in the presence of missing and/or latent variables. The ORF model being considered in this paper can be placed squarely in the original mold for which the method was
developed by treating the latent vectors $\boldsymbol{\xi }_{j}$ as missing. The EM\ algorithm takes the log-likelihood function of the
full data -- observed variables $(Y_{ij},T_{ij})$ and unobserved variables $\boldsymbol{\xi }_{j}$ -- and then iterates between calculating the expected value of the log-likelihood function
conditional on the observed random variables (E-step) and maximizing the function obtained in said step in terms of the model parameters (M-step). This iterative process is repeated until convergence of the model parameters is achieved. In
this subsection, the two steps of the EM algorithm are formalized in the context of the ORF model. Furthermore, a Monte Carlo approach for the E-step is proposed, as closed form expressions are not available for the conditional expectations that need to be calculated.

The complete data likelihood function is%
\begin{eqnarray*}
	\mathcal{L} &\mathcal{=}&\prod_{j=1}^{n}f_{\mathbf{Y},\mathbf{T},\boldsymbol{\xi 
		}|\mathcal{S}_{j}}\left( \mathbf{y}_{j},\mathbf{t}_{j},\theta _{j},\tau _{j}|%
	\mathcal{S}_{j}\right)  \\
	&=&\prod_{j=1}^{n}\left[ \prod_{i\in \mathcal{S}_{j}}f\left(
	y_{ij};N_{i},p_{i}\left( \theta \right) \right) \alpha _{i}\phi \left(
	\alpha _{i}\left( t_{ij} -\beta _{i}+\tau_j \right) \right) \right] \phi _{2}\left( 
	\boldsymbol{\xi }_{j};\mathbf{\mu }_{\xi },\mathbf{\Sigma }_{\xi }\right) 
\end{eqnarray*}%
which can also be written as%
\begin{eqnarray*}
	\mathcal{L}\ &\mathcal{=}\ &\prod_{j=1}^{n}\prod_{i\in \mathcal{S}_{j}}\dbinom{%
		N_{i}}{y_{ij}}\Phi \left[ a_{i}\left( \theta _{j}-b_{i}\right) \right]
	^{y_{ij}}\left\{ 1-\Phi \left[ a_{i}\left( \theta _{j}-b_{i}\right) \right]
	\right\} ^{N_{i}-y_{ij}} \\
	&&\times \prod_{j=1}^{n}\prod_{i\in \mathcal{S}_{j}}\frac{\alpha _{i}}{\sqrt{%
			2\pi }}\exp \left[ -\frac{1}{2}\alpha _{i}^{2}\left( t_{ij}-\beta _{i}+\tau
	_{j}\right) ^{2}\right]  \\
	&&\times \prod_{j=1}^{n}\frac{1}{2\pi \left( \sigma _{\tau }^{2}-\sigma
		_{\theta \tau }^{2}\right) ^{1/2}}\exp \left\{ -\frac{1}{2\left( \sigma
		_{\tau }^{2}-\sigma _{\theta \tau }^{2}\right) }\left( \sigma _{\tau
	}^{2}\theta _{j}^{2}-2\sigma _{\theta \tau }\theta _{j}\tau _{j}+\tau
	_{j}^{2}\right) \right\} .
\end{eqnarray*}%
Let $\mathbf{\Xi }$ denote the collection of all model parameters and let $\mathbf{\hat{\Xi}}_{0}$
denote some initial values for these parameters (possibly the MOM estimators from the previous subsection). Let $\mathbf{\hat{\Xi}}%
_{k}$ denote the parameter estimates obtained after the $k^{th}$ iteration of the EM
algorithm. Define the conditional expectation function%
\begin{equation*}
Q\left( \mathbf{\Xi ,\hat{\Xi}}_{k-1}\right) =E_{\mathbf{\hat{\Xi}}_{k-1}}\left[ \hbox{log}\mathcal{L}\ %
\Bigr|\left( \mathbf{y}_{j},\mathbf{t}_{j},\mathcal{S}_{j}\right) ,j=1,\ldots ,n, \right]
\end{equation*}%
where the conditional expectation is evaluated treating $\mathbf{\hat{\Xi}}_{k-1}$ as the true parameter values. Here, up to a constant of proportionality that does not depend on the
parameter values,%
\begin{eqnarray*}
	\hbox{log}\mathcal{L} &=&\sum_{j=1}^{n}\sum_{i\in \mathcal{S}_{j}}y_{ij}%
	\hbox{log}\Phi \left[ a_{i}\left( \theta _{j}-b_{i}\right) \right]
	+\sum_{j=1}^{n}\sum_{i\in \mathcal{S}_{j}}\left( N_{i}-y_{ij}\right) %
	\hbox{log}\left\{ 1-\Phi \left[ a_{i}\left( \theta _{j}-b_{i}\right) \right]
	\right\}  \\
	&&+\sum_{j=1}^{n} \sum_{i\in \mathcal{S}_{j}}\hbox{log}\alpha _{i}-\frac{1}{2}%
	\sum_{j=1}^{n}\sum_{i\in \mathcal{S}_{j}}\alpha _{i}^{2}\left( t_{ij}-\beta
	_{i}+\tau _{j}\right) ^{2}-\frac{n}{2}\log \left( \sigma _{\tau }^{2}-\sigma
	_{\theta \tau }^{2}\right)  \\
	&&-\frac{1}{2\left( \sigma _{\tau }^{2}-\sigma _{\theta \tau }^{2}\right) }%
	\sum_{j=1}^{n}\left( \sigma _{\tau }^{2}\theta _{j}^{2}-2\sigma _{\theta
		\tau }\theta _{j}\tau_{j}+\tau_{j}^{2}\right) 
\end{eqnarray*}%
and thus,%
\begin{eqnarray}
Q\left( \mathbf{\Xi ,\hat{\Xi}}_{k-1}\right)  &=&\sum_{j=1}^{n}\sum_{i\in 
	\mathcal{S}_{j}}y_{ij}E_{\mathbf{\hat{\Xi}}_{k-1}}\left[ \hbox{log}\Phi \left[ a_{i}\left( \theta
_{j}-b_{i}\right) \right] \Bigr|\mathbf{y}_{j},\mathbf{t}_{j},\mathcal{S}_{j}
\right]   \notag \\
&&+\sum_{j=1}^{n}\sum_{i\in \mathcal{S}_{j}}\left( N_{i}-y_{ij}\right) E_{\mathbf{\hat{\Xi}}_{k-1}}
\left[ \hbox{log}\left\{ 1-\Phi \left[ a_{i}\left( \theta _{j}-b_{i}\right) %
\right] \right\} \Bigr|\mathbf{y}_{j},\mathbf{t}_{j},\mathcal{S}_{j}\right]   \notag \\
&&+\sum_{j=1}^{n}\sum_{i\in \mathcal{S}_{j}}\hbox{log}\alpha _{i}-\frac{1}{2}%
\sum_{j=1}^{n}\sum_{i\in \mathcal{S}_{j}}\alpha _{i}^{2}E_{\mathbf{\hat{\Xi}}_{k-1}}\left[ \left(
t_{ij}-\beta _{i}+\tau _{j}\right) ^{2}\Bigr|\mathbf{y}_{j},\mathbf{t}_{j},%
\mathcal{S}_{j}\right] -\frac{n}{2}\log \left(
\sigma _{\tau }^{2}-\sigma _{\theta \tau }^{2}\right)   \notag \\
&&.-\frac{1}{2\left( \sigma _{\tau }^{2}-\sigma _{\theta \tau }^{2}\right) }%
\sum_{j=1}^{n}E_{\mathbf{\hat{\Xi}}_{k-1}}\left[ \sigma _{\tau }^{2}\theta _{j}^{2}-2\sigma _{\theta
	\tau }\theta _{j}\tau _{j}+\tau _{j}^{2}\Bigr|\mathbf{y}_{j},\mathbf{t}_{j},%
\mathcal{S}_{j}\right] .  
\label{Q function}
\end{eqnarray}

At each step of the EM algorithm, the conditional expectation terms in (\ref{Q function}) are evaluated treating $\boldsymbol{\hat{\Xi}}_{k-1}$ as the true parameter values and thereafter new maximizer  $\boldsymbol{\hat{\Xi}}_{k}$ is found. This iterative process is repeated until convergence of the algorithm is achieved. 

For the ORF model under consideration, there are no closed form solutions for the conditional expectations in (\ref{Q function}). However, it is possible
to sample from the conditional distribution $\left[ \bm{\xi}_j \Bigr|%
\mathbf{y}_{j},\mathbf{t}_{j},\mathcal{S}_{j},\bm{\hat{\Xi}}_{k-1}\right]
$. This sampling algorithm is outlined in the appendix. This allows one to
use the MCEM algorithm as described in Wei \& Tanner (1990). Let $\bm{\xi}_j^{\left( m\right) } =( \theta
_{j}^{\left( m\right) },\tau _{j}^{\left( m\right) }) $, $m=1,\ldots
,M_{k}$ denote $M_{k}$ independent draws from $\left[ \boldsymbol{\xi}_{j} \Bigr|\mathbf{y}_{j},\mathbf{t}_{j},\mathcal{S}_{j},\mathbf{\hat{\Xi}}_{k-1}\right]
$, the distribution of the latent vector $\boldsymbol{\xi}_{j}$ conditional on the observed (non-missing) values $(\mathbf{y}_{j},\mathbf{t}_{j})$ and assuming true parameter values $\mathbf{\hat{\Xi}}_{k-1}$. Define the Monte Carlo approximation to (\ref{Q function}),%
\begin{eqnarray}
\hat{Q}_{k-1}\left( \mathbf{\Xi }\right)  &=&\frac{1}{M_{k}}%
\sum_{m=1}^{M_{k}}\sum_{j=1}^{n}\sum_{i\in \mathcal{S}_{j}}y_{ij}\hbox{log}%
\Phi \left[ a_{i}\left( \theta _{j}^{\left( m\right) }-b_{i}\right) \right] \notag \\&& +%
\frac{1}{M_{k}}\sum_{m=1}^{M_{k}}\sum_{j=1}^{n}\sum_{i\in \mathcal{S}%
	_{j}}\left( N_{i}-y_{ij}\right) \hbox{log}\left\{ 1-\Phi \left[ a_{i}\left(
\theta _{j}^{\left( m\right) }-b_{i}\right) \right] \right\} 
\notag \\
&&+\sum_{j=1}^{n}\sum_{i\in \mathcal{S}_{j}}\hbox{log}\alpha _{i}-\frac{1}{2M_{k}}%
\sum_{m=1}^{M_{k}}\sum_{j=1}^{n}\sum_{i\in \mathcal{S}_{j}}\alpha
_{i}^{2}\left( t_{ij}-\beta _{i}+\tau _{j}^{\left( m\right) }\right) ^{2} 
-\frac{n}{2}\log \left( \sigma _{\tau }^{2}-\sigma _{\theta \tau
}^{2}\right) \notag \\
&& -\frac{1}{2\left( \sigma _{\tau }^{2}-\sigma _{\theta \tau
	}^{2}\right) }\frac{1}{M_{k}}\sum_{m=1}^{M_{k}}\sum_{j=1}^{n}\left[ \sigma
_{\tau }^{2}\left( \theta _{j}^{\left( m\right) }\right) ^{2}-2\sigma
_{\theta \tau }\theta _{j}^{\left( m\right) }\tau _{j}^{\left( m\right)
}+\left( \tau _{j}^{\left( m\right) }\right) ^{2}\right] .  \label{Q_hat function}
\end{eqnarray}%

\medskip
At each iteration of the MCEM algorithm, this function is maximized and thereafter the updated parameter estimates are used to generate a new sample to update the function $\hat{Q}$. Due to the stochastic nature of the MCEM algorithm, Wei \& Tanner (1990) propose selecting a small value $M_{k}$ for the first several
iterations. After these iterations, the updated solution is typically in the part of the parameter space "close to" the maximum likelihood solution. Thereafter, a large value of $%
M_{k}$ ensures that the maximum of $\hat{Q}$ is close to the maximum of $Q$.

\section{Simulation Study}

Several simulation studies were performed to compare the method of moments and Monte Carlo EM-based maximum
likelihood estimators. The simulation study is motivated by the knowledge that the MOM estimators can be computed very quickly, while the MCEM estimators are time-consuming to compute. A representative example of one such simulation is presented here.  This paper does not attempt to make a strict recommendation of one type of estimator over the other, but does illustrate the difference between the two methods in terms of RMSE.

For the $i^{th}$ individual, a pair of latent traits $%
\left( \theta _{i},\tau _{i}\right) $ was simulated from a bivariate normal distribution with fixed parameters $\mu_\theta=\mu_\tau=0$ and  $\sigma^2_\theta=1$. For the simulation presented here, $\sigma^2_\tau=0.24155^2$ and $\sigma_{\theta,\tau}=-0.18116$. Note that the variance parameter $\sigma^2_\tau$ is a scale-dependent parameter. That is, the scale of measurement for time, in this simulation taken to be minutes, affects the size of the parameter. The parameter $\sigma_{\theta,\tau}$ was chosen to give a correlation of $\rho=-0.75$ between the two latent traits. These latent trait vectors were then used to simulate data under two different item configurations. Data $\left( \mathbf{Y}_{i}^{\left(
	1\right) },\mathbf{T}_{i}^{\left( 1\right) }\right) $ was generated under a scenario with $I=2$ items each consisting of $N_i=50$ words, while data $\left( \mathbf{Y}_{i}^{\left( 2\right) },\mathbf{T}%
_{i}^{\left( 2\right) }\right) $ was generated under a scenario with $I=4$ items each consisting of $N_i=25$ words. These parameter specifications all correspond approximately to an average oral reading speed of $122.5$ WPM and success rate of $98$ WCPM. Data were generated with sample sizes $n \in \{40,100,250\}$.

\begin{center}
	\dotfill \\
	\bigskip
	Table \ref{Items_summary_info} about here \\
	\bigskip
	\dotfill \\
\end{center}

In Table \ref{Items_summary_info}, the means and variances of the words correct counts $Y_{ij}^{\left( \ell \right) }$ and reading times (not log-scale) $T_{ij}^{\left( \ell \right) }$ are summarized for the two simulation scenarios $\ell=1,2$. Model parameters can easily be recovered from these using the moment equations (\ref{E_Y}) through (\ref{Var_logT}) and are therefore not presented separately. Note that the parameters were chosen so that the total number of words read correctly, $\sum_{i}Y_{ij}^{(\ell)}$, and total reading time, $\sum_{i}T_{ij}^{(\ell)}$, have, conditional on the latent traits, the same means and variances for both scenarios $\ell=1,2$. 

A total of $500$ samples were simulated in this way. For each sample, the MOM and MCEM estimators were calculated. The MCEM algorithm consisted of $K=13$ iterations where the first $10$ iterations used $M=20$ and the last three iterations used $M=200$ Monte Carlo imputations. As the parameters $(a_i,b_i,\alpha_i,\beta_i)$, $i=1,\ldots,I$ were the same across all $I$ items in the simulation specifications, the average standard error (ASE) and average root mean square error (ARMSE), found by calculating the average of the item-specific standard errors and root mean square errors, are reported in Tables \ref{Sim_N25_I4} and \ref{Sim_N50_I2}. The quantities are also scaled by $n^{1/2}$, the asymptotic convergence rate of the parameters.

\begin{center}
	\dotfill \\
	\bigskip
	Tables \ref{Sim_N25_I4} \& \ref{Sim_N50_I2} about here \\
	\bigskip
	\dotfill \\
\end{center}

Inspection of Tables \ref{Sim_N25_I4} and \ref{Sim_N50_I2} reveal several interesting points. Firstly, note that ARMSE is only slightly larger than ASE. On average, for the MOM estimates, ARMSE represents approximately a $1\%$ increase over ASE, while for the ML estimates, ARMSE represents about a $0.7\%$ increase over ASE. This is strongly indicative that both the MOM and ML parameter estimates are nearly unbiased. Furthermore, with the exception of the precision parameters $\alpha_i$, ASE and ARMSE are relatively stable with respect to sample size. However, when the sample size is small, there is great uncertainty in estimating $\alpha_i$, as is evident from the large ASE and ARMSE values at $n=40$. Here, the ML estimators of the precision parameters have substantially lower variability than the MOM estimators. For $\alpha_i$, there is between a $10\%$ and $50\%$ reduction in ARMSE when comparing MOM and MLE estimators. The effect is not as pronounced when considering the other parameters, with the ARMSE of both $a_i$ and $b_i$ decreasing, but that of $\beta_i$ slightly increasing when comparing the MOM and ML estimators.

Inspecting the performance of the two methods of estimation is interesting, but a more relevant question is the ability of the model to recover the latent traits of individuals. This was also investigated in the simulation study. For each simulated dataset, both the MOM and ML parameter estimates were used to estimate individual traits $\hat{\theta}_{j}=E\left[ \theta _{j}|%
\mathbf{Y}_{j},\mathbf{T}_{j}\right] $ and $\hat{\tau}_{j}=E\left[ \tau _{j}|%
\mathbf{Y}_{j},\mathbf{T}_{j}\right] $ by taking $M=20$ Monte Carlo draws from the conditional distribution and averaging these. In each instance, the correlation coefficients between the imputed latent variables and the known true values in the simulated data was computed. The average correlations across the $500$ simulated datasets are reported in Tables \ref{Corr_N25_I4} and \ref{Corr_N50_I2}.

\begin{center}
	\dotfill \\
	\bigskip
	Tables \ref{Corr_N25_I4} \& \ref{Corr_N50_I2} about here \\
	\bigskip
	\dotfill \\
\end{center}

The results in Tables \ref{Corr_N25_I4} and \ref{Corr_N50_I2} are very encouraging. Even at a small sample size of $n=40$, the latent traits are recovered well as indicated by large average correlations. The model using $I=4$ items with $N=25$ words per item does a much better job of recovering latent traits than the model with only $I=2$ items with $N=50$ words each. This is interesting, since the means and variances of the two settings are the same conditional on the latent traits. Also note that for the parameter specifications used, reading accuracy ability $\theta$ is recovered with greater precision than reading speed ability $\tau$. Overall, increasing the sample size results in a larger average correlation, and maximum likelihood performs better at the same sample size than method of moments does.

\section{Data Analysis}

The data analyzed in this section were part of the data collected from two public school districts in the Pacific Northwest region in the United States. The data consists $n=53$ fourth graders, and measurements $%
\left( \mathbf{Y}_{j},\log \mathbf{T}_{j}\right) $ were obtained for $I=18$
items at the sentence level.  Sentences vary in length, being as short as
just four words and as long as nineteen words. The average proportion of
words read correctly for these items is in the range $0.828$ to $0.957$. For
each item, there are a few missing values. This indicates that a specific
student did not read a paragraph which contained said items. The
item-specific missingness rate ranges from $0.094$ to $0.151$. For a given
item, a few missing values are not a problem, but there are only $35$
complete data cases (approximately $66\%$ of the sample). With an assumption
that values are missing completely at random, the MCEM
algorithm adjusting for missing values will be used.

\begin{center}
	\dotfill \\
	\bigskip
	Table \ref{Data_Analysis} about here \\
	\bigskip
	\dotfill \\
\end{center}

Table \ref{Data_Analysis} provides a summary of the data, listing the number of words in each item, the observed sample means
and standard deviations calculated by ignoring missing values, as well as the means and standard deviations implied by the model after implementing the MCEM algorithm. The model-implied moments are a one-to-one transformation of the estimated model parameters along with variance estimate $\hat{\sigma}_{\tau}^2 = 0.0469$. The estimated correlation between the two latent variables is $\hat{\rho} = -0.037 $. The tabulated model-implied moments conform closely to the sample moments shown and and are suggestive that the model has a large amount of agreement with the data. There is presently no formal goodness-of-fit test for this setting and is a future avenue of research.

Also presented in this section is an assessment of the predictive power of
this model. To mimic the idea of out-of-sample predictive ability, the
analysis was done as follows. Let $\mathbf{\hat{\Xi}}$ denote the maximum
likelihood estimators of the model parameters. Let $\mathbf{y}_{\left(
	-i\right) j}$ and $\mathbf{t}_{\left( -i\right) j}$ denote the count and
log-time vectors for the $j^{th}$ individual with the $i^{th}$ item removed.
Also let $\mathcal{S}_{\left( -i\right) j}=\mathcal{S}_{j}\backslash \left\{
i\right\} $. Let $\left( \theta _{\left( -i\right) j}^{\left( m\right)
},\tau _{\left( -i\right) j}^{\left( m\right) }\right) $, $m=1,\ldots ,M$
denote $M$ pairs sampled from $\left[ \theta _{j},\tau _{j}|\mathbf{y}%
_{\left( -i\right) j},\mathbf{t}_{\left( -i\right) j},\mathcal{S}_{\left(
	-i\right) j},\mathbf{\hat{\Xi}}\right] $. The estimated latent scores%
\begin{equation*}
\hat{\theta}_{\left( -i\right) j}=\frac{1}{M}\sum_{m=1}^{M}\theta _{\left(
	-i\right) j}^{\left( m\right) }
\end{equation*}%
and%
\begin{equation*}
\hat{\tau}_{\left( -i\right) j}=\frac{1}{M}\sum_{m=1}^{M}\theta _{\left(
	-i\right) j}^{\left( m\right) }
\end{equation*}%
are therefore estimates of the latent traits obtained without using information from the $i^{th}$ item.
Noting that $E\left[ Y_{ij}\right] =n_{i}\Phi \left( -a_{i}b_{i}/\sqrt{%
	1+a_{i}^{2}}\right) $ and $E\left[ Y_{ij}|\theta _{j}\right] =n_{i}\Phi %
\left[ a_{i}\left( \theta _{j}-b_{i}\right) \right] $, reasonable predictors
of $Y_{ij}$ are $\hat{Y}_{ij}^{\left( 0\right) }=n_{i}\Phi \left( -\hat{a}%
_{i}\hat{b}_{i}/\sqrt{1+\hat{a}_{i}^{2}}\right) $ and $Y_{ij}^{\left(
	1\right) }=n_{i}\Phi \left[ \hat{a}_{i}\left( \hat{\theta}_{\left( -i\right)
	j}-\hat{b}_{i}\right) \right] $ where $\hat{Y}_{ij}^{\left( 0\right) }$ is
the best predictor without any information on the latent factor, and $\hat{Y}%
_{ij}^{\left( 1\right) }$ is the predictor using the latent factor estimated
without using information from the $i^{th}$ item. Similarly, noting that $E%
\left[ T_{ij}\right] =\exp \left[ \beta _{i}+\sigma _{\tau }^{2}/2+1/\left(
2\alpha _{i}^{2}\right) \right] $ and $E\left[ T_{ij}|\tau _{j}\right] =\exp %
\left[ \beta _{i}-\tau _{j}+1/\left( 2\alpha _{i}^{2}\right) \right] $,
predictors $\hat{T}_{ij}^{\left( 0\right) }=\exp \left[ \hat{\beta}_{i}+\hat{%
	\sigma}_{\tau }^{2}/2+1/\left( 2\hat{\alpha}_{i}^{2}\right) \right] $ and%
\newline
$\hat{T}_{ij}^{\left( 1\right) }=\exp \left[ \hat{\beta}_{i}-\hat{\tau}%
_{\left( -i\right) j}+1/\left( 2\hat{\alpha}_{i}^{2}\right) \right] $ are
defined.

Define the root square prediction error, $RSPE_{Y_{i}}^{\left( k\right) }=%
\left[ \left\vert S_{i}\right\vert ^{-1}\sum_{j\in S_{i}}\left( Y_{ij}-\hat{Y%
}_{ij}^{\left( k\right) }\right) ^{2}\right] ^{1/2}$ and $%
RSPE_{T_{i}}^{\left( k\right) }=\left[ \left\vert S_{i}\right\vert
^{-1}\sum_{j\in S_{i}}\left( T_{ij}-\hat{T}_{ij}^{\left( k\right) }\right)
^{2}\right] ^{1/2}$ for $k=0,1$. These values are reported in Table \ref{Prediction_Error}.

\pagebreak

\begin{center}
	\dotfill \\
	\bigskip
	Table \ref{Prediction_Error} about here \\
	\bigskip
	\dotfill \\
\end{center}

In general, the RSPE\ values in Table \ref{Prediction_Error} indicate that the latent factors have strong predictive power. When considering the relative decrease in
RSPE, $\left( RSPE^{\left( 0\right) }-RSPE^{\left( 1\right) }\right)
/RSPE^{\left( 0\right) }$, these values range between $0.097$ and $0.454$
for the count data, with an average reduction in RSPE of about $24\%$. For
the time data, the values range between $0.122$ and $0.507$, with an average
reduction in RSPE of about $30\%$.

\section{Conclusion}

A latent joint model to measure oral reading speed and accuracy was proposed in this paper and a Monte Carlo EM algorithm to estimate the model parameters was derived. Overall, it was demonstrated that model parameters were estimated well. Simulation studies demonstrated reasonable recovery of model parameters. Also, data analysis demonstrated the latent factors had excellent predictive power. However, the quality of estimated speed factor scores as reporting oral reading speed is still unknown. Therefore, future research to investigate the characteristics of estimated speed factor scores is warranted.

\pagebreak

\section{Appendix}

\subsection{Derivation of Some Model Moments}

The derivation of the moments for the ORF model outlined in this paper relies on two important results concerning the normal distribution. The first of these two results is given here before the derivation the moments of $Y_{ij}$, while the second result is presented before the derivation of the covariance between $Y_{ij}$ and $\log T_{ij}$ is shown.

\textbf{Result 1}: \textit{Let $Z$ be standard normal random variable and let $a>0$ and $b$ be real numbers. It is then true that $\mathrm{E} \left[ \Phi \left( a\left( Z -b\right) \right) \right]=\Phi \left[ -ab / (1+a^2)^{1/2}\right]$ and $\mathrm{E} \left[ \Phi^2 \left( a\left( Z -b\right) \right) \right]=\Phi_2 \left[ -ab / (1+a^2)^{1/2},-ab/(1+a^2)^{1/2}|\rho=a^2/(1+a^2)\right]$ where $\Phi(\cdot)$ denotes the standard normal cdf and $\Phi_2(\cdot,\cdot|\rho)$ denotes the bivariate normal cdf with zero means, unit variances and correlation $\rho$.}

\textbf{Proof}: Let $Z_1,Z_2$ and $Z_3$ be \textit{iid} standard normal random variables. Then,
\begin{eqnarray*}
	E\left[ \Phi \left( a\left( Z_2 -b\right) \right) \right] &=&P\left(
	Z_1<a\left( Z_2 -b\right) \right) \\
	&=&P\left( (1+a^2)^{1/2} Z_1<-ab\right) \\
	&=&\Phi \left[ -ab/(1+a^2)^{1/2}\right]
\end{eqnarray*}
where the second line follows from noting that a linear combination of independent normal variables is again normal. Also,
\begin{eqnarray*}
	E\left[ \Phi ^{2}\left( a\left( Z_1-b\right) \right) \right] &=&\int_{\mathbb{R}}\left(
	\int_{-\infty }^{a\left( z-b\right) }\phi \left( t\right) dt\cdot
	\int_{-\infty }^{a\left( z-b\right) }\phi \left( s\right) ds\right) \phi
	\left( z\right) dz \\
	&=&\int_{\mathbb{R}}\left( \int_{\mathbb{R}}\int_{\mathbb{R}}I\left( t\leq a\left( z-b\right) \right)
	I\left( s\leq a\left( z-b\right) \right) \phi \left( t\right) \phi \left(
	s\right) dtds\right) \phi \left( z\right) dz \\
	&=&\int_{\mathbb{R}}P\left( Z_2\leq a\left( z-b\right) ,Z_3\leq a\left( z-b\right)
	\right) \phi \left( z\right) dz \\
	&=&P\left( Z_2-aZ_1\leq -ab,Z_3-aZ_1\leq -ab\right) \\
	&=&\Phi_2 \left[ -ab / (1+a^2)^{1/2},-ab/(1+a^2)^{1/2}|\rho=a^2/(1+a^2)\right]
\end{eqnarray*}
where the last equality follows upon noting that the pair $(Z_2-aZ_1,Z_3-aZ_1)$ is bivariate normal with correlation $\rho=a^2/(1+a^2)$. 

\bigskip

For random variable $Y_{ij}$, we have by application of the tower property of expectation,
\[
\mathrm{E} \left[ Y_{ij} \right] = \mathrm{E} \left[\mathrm{E} (Y_{ij}|\theta_j) \right] = N_i \mathrm{E} \left[  \Phi (a_i(\theta_j-b_i)\right]
\]
and similarly, 
\[
\mathrm{E} \left[ Y_{ij}^2 \right] = \mathrm{E} \left[\mathrm{E} (Y_{ij}^2|\theta_j) \right]
= N_i(N_i-1)\mathrm{E} \left[  \Phi^2 (a_i(\theta_j-b_i)\right] + N_i \mathrm{E} \left[  \Phi (a_i(\theta_j-b_i)\right]
\]
from which
\begin{eqnarray*}
	\mathrm{Var}\left( Y_{ij}\right) &=&N_{i}^{2}\left[ \Phi _{2}\left( -\frac{a_{i}b_{i}}{(1+a_{i}^{2})^{1/2}},-%
	\frac{a_{i}b_{i}}{(1+a_{i}^{2})^{1/2}}\left\vert \rho =\frac{a_{i}^{2}}{%
		1+a_{i}^{2}}\right. \right) -\Phi ^{2}\left( -\frac{a_{i}b_{i}}{(1+a_{i}^{2})^{1/2}}\right) \right] \\
	&&+N_{i}\left[ \Phi \left( -\frac{a_{i}b_{i}}{(1+a_{i}^{2})^{1/2}}\right)
	-\Phi _{2}\left( -\frac{a_{i}b_{i}}{(1+a_{i}^{2})^{1/2}},-\frac{a_{i}b_{i}}{%
		(1+a_{i}^{2})^{1/2}}\left\vert \rho =\frac{a_{i}^{2}}{1+a_{i}^{2}}\right.
	\right) \right].
\end{eqnarray*}

The result required for calculation of the covariance between $Y_{ij}$ and $\log T_{ij}$ is now presented.

\textbf{Result 2}: \textit{Let $Z$ be standard normal random variable and let $a>0$ and $b$ be real numbers. It is then true that 
	\[
	\mathrm{E} \left[ Z \Phi \left( a\left( Z -b\right) \right) \right]=\mathrm{E} \left[ \phi (b+a^{-1}Z) \right]
	= (2\pi)^{-1/2} \left(\frac{a^2}{1+a^2}\right)^{1/2} \exp\left(-\frac{1}{2} \frac{a^2 b^2}{1+a^2}\right)
	\]
	where $\phi(\cdot)$ denotes the standard normal pdf.}

\textbf{Proof}: The first equality relies on the result $\mathrm{E} \left[ZI(Z\ge x)\right]=\phi(x)$ as well as application of the tower property. The second equality follows from some tedious but straightforward algebra. 

In evaluating the covariance, consider
\begin{eqnarray*}
	\mathrm{E}\left[ Y_{ij}\hbox{log}T_{ij}\right] &=& \mathrm{E}\left[ \mathrm{E}\left[ Y_{ij}|\theta _{j}\right] \mathrm{E}\left[ \hbox{log}T_{ij}|\tau_j \right] \right] \\
	&=& N_i \mathrm{E} \left[\Phi(a_i(\theta_j-b_i))(\beta_i-\tau_j)\right] \\
	&=& N_i \beta_i \mathrm{E} \left[\Phi(a_i(\theta_j-b_i))\right] - N_i \mathrm{E} \left[\tau_j \Phi(a_i(\theta_j-b_i))\right] \\
	&=& N_i \beta_i \mathrm{E} \left[\Phi(a_i(\theta_j-b_i))\right] - N_i \mathrm{E} \left[(\sigma_{\tau \theta}\theta_j + (\sigma_{\tau}^2  - \sigma_{\tau \theta}^2 )^{1/2}Z_j) \Phi(a_i(\theta_j-b_i))\right] \\
	&=& N_i \beta_i \mathrm{E} \left[\Phi(a_i(\theta_j-b_i))\right] - N_i \sigma_{\tau \theta} \mathrm{E} \left[\theta_j \Phi(a_i(\theta_j-b_i))\right] \\
\end{eqnarray*}
where the second-to-last equality makes use of the fact that $\theta_j$ has the same distribution as $\sigma_{\tau \theta}\theta_j + (\sigma_{\tau}^2  - \sigma_{\tau \theta}^2 )^{1/2}Z_j$ where $Z_j$ is a standard normal random variable and is independent of $\theta_j$. It then follows from application of Result 2 that the covariance is given by
\begin{eqnarray*}
	Cov\left( Y_{ij},\hbox{log}T_{ij}\right) &=&E\left[ Y_{ij}\hbox{log}T_{ij}%
	\right] -E\left[ Y_{ij}\right] E\left[ \hbox{log}T_{ij}\right] \\
	&=&-N_{i}\sigma _{\tau \theta }E\left[ \theta _{j}\Phi \left( a_{i}\left(
	\theta _{j}-b_{i}\right) \right) \right] \\
	&=&-\sigma _{\tau \theta }\frac{N_{i}}{\sqrt{2\pi }}\left( \frac{a_{i}^{2}}{%
		a_{i}^{2}+1}\right) ^{1/2}\exp \left( -\frac{1}{2}\frac{a_{i}^{2}b_{i}^{2}}{%
		a_{i}^{2}+1}\right).
\end{eqnarray*}%

\subsection{Sampling Algorithm for MCEM Implementation}
For the $j^{th}$ individual, the joint distribution for the vectors of non-missing counts and log-times, $\mathbf{Y}%
_{j}$ and $\mathbf{T}_{j}$ conditional on the latent variables $\theta _{j}$
and $\tau _{j}$ is
\begin{eqnarray*}
	f_{\mathbf{Y}_{j},\log\mathbf{T}_{j}|\theta _{j},\tau _{j}}\left( \mathbf{y}_{j},%
	\mathbf{t}_{j}|\theta _{j},\tau _{j}\right) &=& \prod_{i\in\mathcal{S}_j}\dbinom{n_{i}}{%
		y_{ij}}\Phi \left[ a_{i}\left( \theta _{j}-b_{i}\right) \right]
	^{y_{ij}}\left\{ 1-\Phi \left[ a_{i}\left( \theta _{j}-b_{i}\right) \right]
	\right\} ^{n-y_{ij}} \\
	&& \times \prod_{i\in\mathcal{S}_j}\frac{\alpha _{i}}{\sqrt{2\pi }}\exp \left[ -\frac{1%
	}{2}\alpha _{i}^{2}\left( t_{ij}-\beta _{i}+\tau _{j}\right) ^{2}\right] 
\end{eqnarray*}
while the latent traits have distribution
\begin{equation*}
f_{\theta _{j},\tau _{j}}\left( \theta _{j},\tau _{j}\right) = \frac{1}{2\pi \left( \sigma _{\tau }^{2}-\sigma _{\theta \tau
	}^{2}\right) ^{1/2}}\exp \left\{ -\frac{1}{2\left( \sigma _{\tau
	}^{2}-\sigma _{\theta \tau }^{2}\right) }\left( \sigma _{\tau }^{2}\theta
_{j}^{2}-2\sigma _{\theta \tau }\theta _{j}\tau _{j}+\tau _{j}^{2}\right)
\right\}.
\end{equation*}
The unconditional distribution of $\left[\bm{Y}_j,\log\bm{T}_j\right]$ involves a double integral and is therefore impractical for direct use, it is possible to integrate out the latent component $\tau_j$,
\[
f_{\mathbf{Y}_{j},\log\mathbf{T}_{j},\theta _{j}}\left( \mathbf{y}_{j},\mathbf{t}%
_{j},\theta _{j}\right) =\int_{\mathbb{R}}f_{\mathbf{Y}_{j},\log\mathbf{T}_{j}|\theta _{j},\tau _{j}}\left( \mathbf{y}_{j},%
\mathbf{t}_{j}|\theta _{j},\tau _{j}\right) f_{\theta _{j},\tau _{j}}\left( \theta _{j},\tau _{j}\right) d\tau _{j} 
\]
and as $\left[\theta_j |\mathbf{Y}_{j},\mathbf{T}_{j},\mathcal{S}_j\right]$ is proportional to $\left[\mathbf{Y}_{j},\mathbf{T}_{j},\theta_j,\mathcal{S}_j\right]$, one can show that
\begin{eqnarray}
f_{\theta _{j} | \mathbf{Y}_{j},\log\mathbf{T}_{j}}\left( \mathbf{y}_{j},\mathbf{t}%
_{j},\theta _{j}\right) &\propto& \left[ \prod_{i\in\mathbb{S}_j}\alpha _{i}%
\dbinom{n_{i}}{y_{ij}}\Phi \left[ a_{i}\left( \theta_{j}-b_{i}\right) \right]
^{y_{ij}}\left\{ 1-\Phi \left[ a_{i}\left( \theta_{j}-b_{i}\right) \right]
\right\} ^{n_{i}-y_{ij}}\right] \label{cond_dist_theta} \\
&\times& \exp\left\{-\frac{1}{2} \left(\frac{1+\sigma_{\tau}^2 A_j}{1+A_j\left(\sigma_{\tau}^2 - \sigma_{\theta,\tau}^2\right)} \right)\left(\theta_j + \left(\frac{\sigma_{\theta,\tau}}{1+A_j \sigma_{\tau}^2}\right)\left(\sum_{i\in\mathcal{S}_j}\alpha_j^2(t_{ij}-\beta_j)\right)\right)^2\right\} \notag
\end{eqnarray}
where $A_j = \sum_{i\in\mathcal{S}_j} \alpha_i$. One can therefore use rejection sampling scheme to draw samples from $\left[\theta_j |\mathbf{Y}_{j},\mathbf{T}_{j},\mathcal{S}_j\right]$. Let $f^{*}\left(\theta_j\right)$ denote the right-hand side of (\ref{cond_dist_theta}) and define
\[
g^{*}(\theta_j) = \exp\left\{-\frac{1}{2} \left(\frac{1+\sigma_{\tau}^2 A_j}{1+A_j\left(\sigma_{\tau}^2 - \sigma_{\theta,\tau}^2\right)} \right)\left(\theta_j + \left(\frac{\sigma_{\theta,\tau}}{1+A_j \sigma_{\tau}^2}\right)\left(\sum_{i\in\mathcal{S}_j}\alpha_j^2(t_{ij}-\beta_j)\right)\right)^2\right\}.
\]
Note that $g^{*}(\theta_j)$ is the kernel of a Gaussian distribution with mean
\begin{equation}
\mu_g = -\left(\frac{\sigma_{\theta,\tau}}{1+A_j \sigma_{\tau}^2}\right)\left(\sum_{i\in\mathcal{S}_j}\alpha_j^2(t_{ij}-\beta_j)\right) \label{mu_g} 
\end{equation}
and variance
\begin{equation}
\sigma_g^2 = \left(\frac{1+\sigma_{\tau}^2 A_j}{1+A_j\left(\sigma_{\tau}^2 - \sigma_{\theta,\tau}^2\right)}\right)^{-1}. \label{sig2_g} 
\end{equation}
Noting that $f^{*}\left(\theta\right)\leq g^{*}\left(\theta\right)$ for all $\theta$ and defining
\[
\gamma^{-1} = \sup_{\theta} \frac{f^{*}(\theta)}{g^{*}(\theta)}
\]
where
\[
\frac{f^{*}(\theta)}{g^{*}(\theta)} = \prod_{i\in\mathbb{S}_j}\alpha _{i} \dbinom{n_{i}}{y_{ij}}\Phi \left[ a_{i}\left( \theta_{j}-b_{i}\right) \right]	^{y_{ij}}\left\{ 1-\Phi \left[ a_{i}\left( \theta_{j}-b_{i}\right) \right]
\right\} ^{n_{i}-y_{ij}}.
\]
Therefore, one can implement rejection sampling in the following way: Generate a normal random number $R$ with mean $\mu_g$ as in (\ref{mu_g}) and variance $\sigma_g^2$ as in (\ref{sig2_g}). Also generate $U$ uniform(0,1). If
\[
U \leq \gamma \frac{f^{*}(R)}{g^{*}(R)}
\]
set $\tilde{\theta}_j=R$, otherwise repeat. Here,  $\tilde{\theta}_j$ represents a random draw from the distribution of $[\theta|\bm{Y},\log \bm{T}]$.

Next, it is easy to verify that the distribution of $[\tau|\theta,\bm{Y},\log \bm{T}]$ is normal with mean
\begin{equation}
\mu_{\tau|\theta,\bm{Y}, \bm{T}} = -\frac{(\sigma_{\tau}^2 - \sigma_{\theta\tau}^2)^{1/2}}{1+A(\sigma_{\tau}^2 - \sigma_{\theta\tau}^2)} \sum_{i\in \mathcal{S}} \alpha_i^2 \left(t_{ij}-\beta_i+\sigma_{\theta\tau}\theta\right) \label{mu_tau}
\end{equation}
and variance
\begin{equation}
\sigma_{\tau|\theta,\bm{Y}, \bm{T}}^2 = \left(1+A\left(\sigma_{\tau}^2 - \sigma_{\theta\tau}^2\right)\right)^{-1}.\label{sig2_tau}
\end{equation}
Therefore, after using rejection sampling to sample $\tilde{\theta}$, the corresponding value $\tilde{\tau}$ can be sampled by plugging $\tilde{\theta}$ into equations (\ref{mu_tau}) and (\ref{sig2_tau}) and generating a normal random variable with that mean and variance.
\pagebreak
\section*{Tables and Figures}


\begin{table}[h]
	\begin{center}
		\begin{tabular}{|c|c|c|}
			\hline
			& $50$ word items & $25$ word items \\ \hline
			$\left( \mu _{count},\sigma _{count}^{2}\right) $ & $\left(
			40,6.2749^{2}\right) $ & $\left( 20,4.4371^{2}\right) $ \\ \hline
			$\left( \mu _{time},\sigma _{time}^{2}\right) $ & $\left(
			0.4086,0.1205^{2}\right) $ & $\left( 0.2043,0.0602^{2}\right) $ \\ \hline
		\end{tabular}%
	\end{center}
	\caption{Means and variances of words correct count and reading time for items under two different simulation scenarios.}
	\label{Items_summary_info}
\end{table}

\begin{table}
	\begin{center}
		
		\begin{tabular}{cc|c|c|c|c|}
			\cline{3-6}
			&  & \multicolumn{2}{|c|}{MOM} & \multicolumn{2}{|c|}{ML} \\ \hline
			\multicolumn{1}{|c}{Sample size} & \multicolumn{1}{|c|}{Parameters} & $%
			n^{1/2}$ASE & $n^{1/2}$ARMSE & $n^{1/2}$ASE & $n^{1/2}$ARMSE \\ \hline
			\multicolumn{1}{|c}{} & \multicolumn{1}{|c|}{$a$} & $0.670$ & $0.704$ & $%
			0.625$ & $0.628$ \\ \cline{2-6}
			\multicolumn{1}{|c}{} & \multicolumn{1}{|c|}{$b$} & $2.123$ & $2.164$ & $%
			1.983$ & $2.011$ \\ \cline{2-6}
			\multicolumn{1}{|c}{$n=40$} & \multicolumn{1}{|c|}{$\alpha $} & $31.813$ & $%
			32.514$ & $21.446$ & $21.648$ \\ \cline{2-6}
			\multicolumn{1}{|c}{} & \multicolumn{1}{|c|}{$\beta $} & $0.285$ & $0.285$ & 
			$0.287$ & $0.297$ \\ \cline{2-6}
			\multicolumn{1}{|c}{} & \multicolumn{1}{|c|}{$\sigma _{\tau }^{2}$} & $0.096$
			& $0.096$ & $0.095$ & $0.095$ \\ \cline{2-6}
			\multicolumn{1}{|c}{} & \multicolumn{1}{|c|}{$\sigma _{\theta \tau }$} & $%
			0.243$ & $0.245$ & $0.228$ & $0.231$ \\ \hline\hline
			\multicolumn{1}{|c}{} & \multicolumn{1}{|c|}{$a$} & $0.696$ & $0.704$ & $%
			0.619$ & $0.625$ \\ \cline{2-6}
			\multicolumn{1}{|c}{} & \multicolumn{1}{|c|}{$b$} & $1.909$ & $1.938$ & $%
			1.825$ & $1.849$ \\ \cline{2-6}
			\multicolumn{1}{|c}{$n=100$} & \multicolumn{1}{|c|}{$\alpha $} & $13.418$ & $%
			13.686$ & $6.009$ & $6.118$ \\ \cline{2-6}
			\multicolumn{1}{|c}{} & \multicolumn{1}{|c|}{$\beta $} & $0.287$ & $0.287$ & 
			$0.287$ & $0.288$ \\ \cline{2-6}
			\multicolumn{1}{|c}{} & \multicolumn{1}{|c|}{$\sigma _{\tau }^{2}$} & $0.092$
			& $0.092$ & $0.091$ & $0.092$ \\ \cline{2-6}
			\multicolumn{1}{|c}{} & \multicolumn{1}{|c|}{$\sigma _{\theta \tau }$} & $%
			0.227$ & $0.227$ & $0.208$ & $0.219$ \\ \hline\hline
			\multicolumn{1}{|c}{} & \multicolumn{1}{|c|}{$a$} & $0.701$ & $0.703$ & $%
			0.601$ & $0.607$ \\ \cline{2-6}
			\multicolumn{1}{|c}{} & \multicolumn{1}{|c|}{$b$} & $1.819$ & $1.823$ & $%
			1.727$ & $1.734$ \\ \cline{2-6}
			\multicolumn{1}{|c}{$n=250$} & \multicolumn{1}{|c|}{$\alpha $} & $9.871$ & $%
			9.969$ & $5.842$ & $5.898$ \\ \cline{2-6}
			\multicolumn{1}{|c}{} & \multicolumn{1}{|c|}{$\beta $} & $0.275$ & $0.276$ & 
			$0.278$ & $0.279$ \\ \cline{2-6}
			\multicolumn{1}{|c}{} & \multicolumn{1}{|c|}{$\sigma _{\tau }^{2}$} & $0.092$
			& $0.093$ & $0.092$ & $0.093$ \\ \cline{2-6}
			\multicolumn{1}{|c}{} & \multicolumn{1}{|c|}{$\sigma _{\theta \tau }$} & $%
			0.232$ & $0.233$ & $0.218$ & $0.235$ \\ \hline
		\end{tabular}%
		\caption{Average standard error (ASE) and average root mean square error (ARMSE) of MOM and EM estimators, $N_{i}=25$ and $I=4$.}
		\label{Sim_N25_I4}
	\end{center}
\end{table}

\begin{table}
	\begin{center}
		
		\begin{tabular}{cc|c|c|c|c|}
			\cline{3-6}
			&  & \multicolumn{2}{|c|}{MOM} & \multicolumn{2}{|c|}{ML} \\ \hline
			\multicolumn{1}{|c}{Sample size} & \multicolumn{1}{|c|}{Parameters} & $%
			n^{1/2}$ASE & $n^{1/2}$ARMSE & $n^{1/2}$ASE & $n^{1/2}$ARMSE \\ \hline
			\multicolumn{1}{|c}{} & \multicolumn{1}{|c|}{$a$} & $0.432$ & $0.434$ & $%
			0.406$ & $0.414$ \\ \cline{2-6}
			\multicolumn{1}{|c}{} & \multicolumn{1}{|c|}{$b$} & $2.709$ & $2.822$ & $%
			2.611$ & $2.716$ \\ \cline{2-6}
			\multicolumn{1}{|c}{$n=40$} & \multicolumn{1}{|c|}{$\alpha $} & $18.606$ & $%
			19.196$ & $15.205$ & $15.937$ \\ \cline{2-6}
			\multicolumn{1}{|c}{} & \multicolumn{1}{|c|}{$\beta $} & $0.288$ & $0.288$ & 
			$0.292$ & $0.292$ \\ \cline{2-6}
			\multicolumn{1}{|c}{} & \multicolumn{1}{|c|}{$\sigma _{\tau }^{2}$} & $0.103$
			& $0.104$ & $0.099$ & $0.100$ \\ \cline{2-6}
			\multicolumn{1}{|c}{} & \multicolumn{1}{|c|}{$\sigma _{\theta \tau }$} & $%
			0.256$ & $0.256$ & $0.231$ & $0.261$ \\ \hline\hline
			\multicolumn{1}{|c}{} & \multicolumn{1}{|c|}{$a$} & $0.411$ & $0.414$ & $%
			0.376$ & $0.383$ \\ \cline{2-6}
			\multicolumn{1}{|c}{} & \multicolumn{1}{|c|}{$b$} & $2.356$ & $2.389$ & $%
			2.266$ & $2.326$ \\ \cline{2-6}
			\multicolumn{1}{|c}{$n=100$} & \multicolumn{1}{|c|}{$\alpha $} & $11.196$ & $%
			11.419$ & $9.289$ & $9.695$ \\ \cline{2-6}
			\multicolumn{1}{|c}{} & \multicolumn{1}{|c|}{$\beta $} & $0.295$ & $0.295$ & 
			$0.299$ & $0.299$ \\ \cline{2-6}
			\multicolumn{1}{|c}{} & \multicolumn{1}{|c|}{$\sigma _{\tau }^{2}$} & $0.100$
			& $0.100$ & $0.096$ & $0.097$ \\ \cline{2-6}
			\multicolumn{1}{|c}{} & \multicolumn{1}{|c|}{$\sigma _{\theta \tau }$} & $%
			0.251$ & $0.252$ & $0.225$ & $0.258$ \\ \hline\hline
			\multicolumn{1}{|c}{} & \multicolumn{1}{|c|}{$a$} & $0.428$ & $0.430$ & $%
			0.390$ & $0.405$ \\ \cline{2-6}
			\multicolumn{1}{|c}{} & \multicolumn{1}{|c|}{$b$} & $2.286$ & $2.301$ & $%
			2.212$ & $2.297$ \\ \cline{2-6}
			\multicolumn{1}{|c}{$n=250$} & \multicolumn{1}{|c|}{$\alpha $} & $9.668$ & $%
			9.732$ & $8.162$ & $8.454$ \\ \cline{2-6}
			\multicolumn{1}{|c}{} & \multicolumn{1}{|c|}{$\beta $} & $0.294$ & $0.294$ & 
			$0.295$ & $0.295$ \\ \cline{2-6}
			\multicolumn{1}{|c}{} & \multicolumn{1}{|c|}{$\sigma _{\tau }^{2}$} & $0.103$
			& $0.103$ & $0.101$ & $0.101$ \\ \cline{2-6}
			\multicolumn{1}{|c}{} & \multicolumn{1}{|c|}{$\sigma _{\theta \tau }$} & $%
			0.262$ & $0.263$ & $0.242$ & $0.311$ \\ \hline
		\end{tabular}%
		\caption{Average standard error (ASE) and average root mean square error (ARMSE) of MOM and EM estimators, $N_{i}=50$ and $I=2$.}
		\label{Sim_N50_I2}
	\end{center}
\end{table}

\begin{table}
	\begin{center}
		
		\begin{tabular}{c|c|c|c|c|c|c|}
			\cline{2-7}
			& \multicolumn{3}{|c}{MOM} & \multicolumn{3}{|c|}{ML} \\ \hline
			\multicolumn{1}{|c}{$n$} & \multicolumn{1}{|c|}{$40$} & $100$ & $250$ & $40$ & $100$ & $250$ \\ \hline
			\multicolumn{1}{|c|}{$\widehat{Cor}\left( \theta ,\hat{\theta}\right) $} & 
			0.9653 & $0.9657$ & $0.9667$ & 0.9662 & $0.9661$ & $0.9668$ \\ \hline
			\multicolumn{1}{|c|}{$\widehat{Cor}\left( \tau ,\hat{\tau}\right) $} & 0.9350
			& $0.9465$ & $0.9505$ & 0.9468 & $0.9504$ & $0.9514$ \\ \hline
		\end{tabular}%
		\caption{Average empirical correlation between true latent scores and imputed scores with $M=20$ Monte Carlo draws for model with $N_{i}=25$ and $I=4$.}
		\label{Corr_N25_I4}
	\end{center}
\end{table}

\begin{table}
	\begin{center}
		
		\begin{tabular}{c|c|c|c|c|c|c|}
			\cline{2-7}
			& \multicolumn{3}{|c}{MOM} & \multicolumn{3}{|c|}{ML} \\ \hline
			\multicolumn{1}{|c}{$n$} & \multicolumn{1}{|c|}{$40$} & $100$ & $250$ & $40$ & $100$ & $250$ \\ \hline
			\multicolumn{1}{|c|}{$\widehat{Cor}\left( \theta ,\hat{\theta}\right) $} & $%
			0.9396$ & $0.9437$ & $0.9429$ & $0.9413$ & $0.9438$ & $0.9427$ \\ \hline
			\multicolumn{1}{|c|}{$\widehat{Cor}\left( \tau ,\hat{\tau}\right) $} & $%
			0.8954$ & $0.9087$ & $0.9121$ & $0.9033$ & $0.9116$ & $0.9134$ \\ \hline
		\end{tabular}%
		\caption{Average empirical correlation between true latent scores and imputed scores with $M=20$ Monte Carlo draws for model with $N_{i}=50$ and $I=2$.}
		\label{Corr_N50_I2}
	\end{center}
\end{table}

\begin{table}
	\begin{center}
		\begin{tabular}{cc|c|c|c|c|}
			\cline{3-6}
			&  & \multicolumn{2}{|c|}{Count Data ($Y$)} & \multicolumn{2}{|c|}{Time Data
				($T$)} \\ \hline
			\multicolumn{1}{|c|}{Item} & \multicolumn{1}{|c|}{$N_{i}$} & Sample Moments
			& Model Moments & Sample Moments & Model Moments \\ 
			\multicolumn{1}{|c|}{} & \multicolumn{1}{|c|}{} & $\left( \bar{y}%
			_{i},s_{y_{i}}\right) $ & $\left( \hat{\mu}_{Y_{i}},\hat{\sigma}%
			_{Y_{i}}\right) $ & $\left( \bar{t}_{i},s_{t_{i}}\right) $ & $\left( \hat{\mu%
			}_{T_{i}},\hat{\sigma}_{T_{i}}\right) $ \\ \hline
			\multicolumn{1}{|c|}{$1$} & \multicolumn{1}{|c|}{$19$} & $\left(
			17.830,2.737\right) $ & $\left( 17.673,2.915\right) $ & $\left(
			7.002,1.649\right) $ & $\left( 6.993,1.726\right) $ \\ \hline
			\multicolumn{1}{|c|}{$2$} & \multicolumn{1}{|c|}{$17$} & $\left(
			16.277,1.728\right) $ & $\left( 16.186,1.944\right) $ & $\left(
			7.139,1.817\right) $ & $\left( 7.122,1.769\right) $ \\ \hline
			\multicolumn{1}{|c|}{$3$} & \multicolumn{1}{|c|}{$8$} & $\left(
			7.383,1.226\right) $ & $\left( 7.397,1.263\right) $ & $\left(
			3.281,0.733\right) $ & $\left( 2.369,0.753\right) $ \\ \hline
			\multicolumn{1}{|c|}{$4$} & \multicolumn{1}{|c|}{$7$} & $\left(
			6.383,1.453\right) $ & $\left( 6.352,1.365\right) $ & $\left(
			3.091,0.786\right) $ & $\left( 3.110,0.887\right) $ \\ \hline
			\multicolumn{1}{|c|}{$5$} & \multicolumn{1}{|c|}{$8$} & $\left(
			7.444,1.486\right) $ & $\left( 7.407,1.447\right) $ & $\left(
			2.829,0.938\right) $ & $\left( 2.812,0.967\right) $ \\ \hline
			\multicolumn{1}{|c|}{$6$} & \multicolumn{1}{|c|}{$5$} & $\left(
			4.644,0.981\right) $ & $\left( 4.627,0.939\right) $ & $\left(
			2.141,0.594\right) $ & $\left( 2.128,0.658\right) $ \\ \hline
			\multicolumn{1}{|c|}{$7$} & \multicolumn{1}{|c|}{$11$} & $\left(
			9.978,2.251\right) $ & $\left( 9.991,2.166\right) $ & $\left(
			4.403,1.786\right) $ & $\left( 4.311,1.473\right) $ \\ \hline
			\multicolumn{1}{|c|}{$8$} & \multicolumn{1}{|c|}{$10$} & $\left(
			8.533,2.007\right) $ & $\left( 8.826,1.924\right) $ & $\left(
			5,322,1.895\right) $ & $\left( 5.211,1.797\right) $ \\ \hline
			\multicolumn{1}{|c|}{$9$} & \multicolumn{1}{|c|}{$13$} & $\left(
			11.356,2.978\right) $ & $\left( 11.555,2.694\right) $ & $\left(
			5.120,1.331\right) $ & $\left( 5.135,1.727\right) $ \\ \hline
			\multicolumn{1}{|c|}{$10$} & \multicolumn{1}{|c|}{$9$} & $\left(
			8.133,2.029\right) $ & $\left( 8.180,1.756\right) $ & $\left(
			3.660,1.411\right) $ & $\left( 3.685,1.368\right) $ \\ \hline
			\multicolumn{1}{|c|}{$11$} & \multicolumn{1}{|c|}{$15$} & $\left(
			13.689,2.922\right) $ & $\left( 13.761,2.581\right) $ & $\left(
			5.753,1.572\right) $ & $\left( 5.768,1.510\right) $ \\ \hline
			\multicolumn{1}{|c|}{$12$} & \multicolumn{1}{|c|}{$8$} & $\left(
			7.178,2.026\right) $ & $\left( 7.061,2.075\right) $ & $\left(
			3.235,1.080\right) $ & $\left( 3.226,0.967\right) $ \\ \hline
			\multicolumn{1}{|c|}{$13$} & \multicolumn{1}{|c|}{$15$} & $\left(
			12.778,3.417\right) $ & $\left( 13.132,3.100\right) $ & $\left(
			6.260,2.306\right) $ & $\left( 6.256,2.016\right) $ \\ \hline
			\multicolumn{1}{|c|}{$14$} & \multicolumn{1}{|c|}{$6$} & $\left(
			5.604,0.962\right) $ & $\left( 5.596,0.947\right) $ & $\left(
			2.374,0.796\right) $ & $\left( 2.376,0.737\right) $ \\ \hline
			\multicolumn{1}{|c|}{$15$} & \multicolumn{1}{|c|}{$16$} & $\left(
			14.958,2.287\right) $ & $\left( 14.955,2.176\right) $ & $\left(
			5.331,1.298\right) $ & $\left( 5.368,1.365\right) $ \\ \hline
			\multicolumn{1}{|c|}{$16$} & \multicolumn{1}{|c|}{$9$} & $\left(
			8.396,1.180\right) $ & $\left( 8.447,1.116\right) $ & $\left(
			3.713,1.129\right) $ & $\left( 3.721,1.085\right) $ \\ \hline
			\multicolumn{1}{|c|}{$17$} & \multicolumn{1}{|c|}{$13$} & $\left(
			11.729,2.248\right) $ & $\left( 11.857,2.006\right) $ & $\left(
			4.845,1.344\right) $ & $\left( 4.870,1.371\right) $ \\ \hline
			\multicolumn{1}{|c|}{$18$} & \multicolumn{1}{|c|}{$4$} & $\left(
			3.313,0.776\right) $ & $\left( 3.446,0.852\right) $ & $\left(
			2.761,1.230\right) $ & $\left( 2.763,1.141\right) $ \\ \hline
		\end{tabular}%
	\end{center}
	\caption{ Data summary: observed item moments (MOM) and model-implied item moments from EM algorithm (ML).}
	\label{Data_Analysis}
\end{table}

\begin{table}
	\begin{center}
		\begin{tabular}{c|c|c|c|c|}
			\cline{2-5}
			& \multicolumn{2}{|c|}{Count Data ($Y$)} & \multicolumn{2}{|c|}{Time Data ($%
				T $)} \\ \hline
			\multicolumn{1}{|c|}{Item} & $RSPE^{\left( 0\right) }$ & $RSPE^{\left(
				1\right) }$ & $RSPE^{\left( 0\right) }$ & $RSPE^{\left( 1\right) }$ \\ \hline
			\multicolumn{1}{|c|}{1} & $2.713$ & $1.480$ & $1.631$ & $0.811$ \\ \hline
			\multicolumn{1}{|c|}{2} & $1.712$ & $1.063$ & $1.798$ & $0.886$ \\ \hline
			\multicolumn{1}{|c|}{3} & $1.213$ & $1.032$ & $0.725$ & $0.506$ \\ \hline
			\multicolumn{1}{|c|}{4} & $1.438$ & $1.108$ & $0.777$ & $0.591$ \\ \hline
			\multicolumn{1}{|c|}{5} & $1.470$ & $1.071$ & $0.928$ & $0.720$ \\ \hline
			\multicolumn{1}{|c|}{6} & $0.970$ & $0.718$ & $0.587$ & $0.391$ \\ \hline
			\multicolumn{1}{|c|}{7} & $2.226$ & $1.559$ & $1.768$ & $1.268$ \\ \hline
			\multicolumn{1}{|c|}{8} & $2.006$ & $1.557$ & $1.877$ & $1.254$ \\ \hline
			\multicolumn{1}{|c|}{9} & $2.952$ & $2.414$ & $1.316$ & $1.057$ \\ \hline
			\multicolumn{1}{|c|}{10} & $2.007$ & $1.813$ & $1.396$ & $1.205$ \\ \hline
			\multicolumn{1}{|c|}{11} & $2.890$ & $2.536$ & $1.555$ & $0.889$ \\ \hline
			\multicolumn{1}{|c|}{12} & $2.007$ & $1.520$ & $1.068$ & $0.699$ \\ \hline
			\multicolumn{1}{|c|}{13} & $3.398$ & $2.557$ & $2.281$ & $1.670$ \\ \hline
			\multicolumn{1}{|c|}{14} & $0.952$ & $0.709$ & $0.788$ & $0.592$ \\ \hline
			\multicolumn{1}{|c|}{15} & $2.264$ & $1.628$ & $1.285$ & $0.891$ \\ \hline
			\multicolumn{1}{|c|}{16} & $1.169$ & $0.850$ & $1.117$ & $0.754$ \\ \hline
			\multicolumn{1}{|c|}{17} & $2.228$ & $1.770$ & $1.330$ & $0.872$ \\ \hline
			\multicolumn{1}{|c|}{18} & $0.780$ & $0.665$ & $1.217$ & $1.068$ \\ \hline
		\end{tabular}%
		\caption{Leave-item-out root square prediction error for count and time data. $RSPE^{(0)}$ is model-free, while $RSPE^{(1)}$ is model-specific.} 
		\label{Prediction_Error}
	\end{center}
\end{table}

\begin{thebibliography}{99}
	
	\bibitem{Andrich78} Andrich, D. (1978). A rating formulation for ordered response categories. \textit{Psychometrika}, 43, 561-73.
	
	\bibitem{Dempster77}  Dempster, A. P., Laird, N. M., \& Rubin, D. B. (1977).
	Maximum likelihood from incomplete data via the EM algorithm'.  \textit{%
		Journal of the royal statistical society. Series B (methodological)}, 1-38.
	
	\bibitem{Entink09} Entink, R. H. K, Kuhn, J-T, Hornke, L. F., \& Fox, J-P. (2009). Evaluating cognitive theory: A joint modeling approach using responses and response times. \textit{Psychological Methods}, 14, 54-75.
	
	\bibitem{Fox07} Fox, J-P., Entink, R., \& van der Linden, W. (2007). Modeling of responses and response time with the package CIRT, \textit{Journal of Statistical Software}, 20, 1-14.
	
	\bibitem{Fuchs04} Fuchs, L. S. (2004). The past, present, and future of curriculum-based measurement research. \textit{School Psychology Review}, 33(2), 188-193.
	
	\bibitem{Levine01} Levine, R. A., \& Casella, G. (2001). Implementations of the Monte Carlo EM algorithm. \textit{Journal of Computational and Graphical Statistics} 10(3), 422-439.
	
	\bibitem{McLachlan07} McLachlan, G. \& Thriyambakam, K. (2007). The EM Algorithm and Extensions. John Wiley \& Sons
	
	\bibitem{vanderLinden07}  van der Linden, W. J. (2007). A Hierarchical Framework for Modeling Speed and Accuracy on Test Items. \textit{Psychometrika} 72(3), 287-308.
	
	\bibitem{vanderLinden10} van der Linden, W. J., Entink, R. H. K., \& Fox, J-P. (2010). IRT parameter estimation with response times as collateral information. \textit{Applied Psychological Measurement}, 34, 327-347.
	
	\bibitem{Wei90}  Wei G. C., \& Tanner, M. A.(1990).  A Monte Carlo
	implementation of the EM algorithm and the poor man's data augmentation
	algorithms.  \textit{Journal of the American statistical Association},
	85(411), 699-704.
	
\end{thebibliography}
\end{document}